\documentclass[12pt,draftclsnofoot,onecolumn]{IEEEtran} 
\usepackage{graphicx,cite,enumitem,}             		
			
\usepackage{amssymb,amsfonts,amsthm}   

\usepackage[cmex10]{amsmath}
\begin{document}

\title{{\LARGE Unified Receiver Design in Wireless Relay Networks Using Mixed-Integer Programming Techniques}}

\author{Kun~Wang \\ \today
}

\maketitle

\begin{abstract}
Wireless receiver design is critical to the overall system performance. 
In this work, we apply the techniques of mixed-integer programming to 
formulate a unified receiver in relay networks with only partial channel information.
We also relax all the constraints into real field to save computations,
but at the cost of bit error performance. 
Moreover, a receiver formulation with adaptive number of constraints is proposed
using the cutting plane techniques. 
Last but not least, number of flops is investigated for different formulations.
\end{abstract}

\section{Introduction}
Diversity has been shown to be an effective technique in combating fading and delivering reliable transmission in wireless communications. Specifically, spatial diversity can be achieved with multiple-input multiple-output (MIMO) systems by employing antenna array at the transmitter and/or receiver side \cite{gesbert2003theory}.
However, in cellular networks or wireless sensor networks, the mobile equipments may not be able to support multiple antennas due to size, cost and hardware limitations. In this case, a new form of diversity, called user cooperation diversity or distributed spatial diversity, has been proposed to reap the benefits of MIMO systems \cite{sendonaris2003user1,sendonaris2003user2}.
In such cooperative communication networks, the users share and coordinate their resources to enhance the transmission quality. This idea is particularly attractive in wireless environments due to the diverse channel quality and the limited energy and bandwidth resources. With the cooperations, users that experience a deep fade can make use of the quality channels provided by their partners to achieve the desired quality of service (QoS) \cite{nosratinia2004cooperative,li2009distributed,hong2007cooperative}.

Typically, the overall transmission in cooperative relay networks can be divided into two phases -- coordination phase and cooperative transmission phase, where the coordination is mainly used for synchronization, node selection and so on \cite{hong2007cooperative}. During the transmission phase, the source node first broadcasts its messages to both relays and destination. Upon receiving signals from the source, each relay node processes the received signals and then forwards them to the destination.
Usually, the relay follows a certain protocol, including demodulation/decode-and-forward \cite{sendonaris2003user1}, amplify-and-forward \cite{laneman2004cooperative}, coded-cooperation \cite{janani2004coded}, and compress-and-forward \cite{kramer2005cooperative}. Different protocols have their advantages as well as disadvantages. In the following, we employ the decode-and-forward (DF) protocol, in which the relay decodes the received signal and re-encodes them before forwarding. Obviously, this scheme avoids noise propagation.
Furthermore, the source and relays can transmit through orthogonal or non-orthogonal channels, in which the orthogonality can be attained by time division, frequency division or code division. In this work, the orthogonal channel is employed in the form of time division by assuming that the source remains silent when relay nodes are transmitting.

Generally, the research topics under the framework of relay network can categorized into node selection \cite{danaee2012relay}, power allocation \cite{pham2009power}, relay protocol design \cite{sendonaris2003user1,wang2007high}, distributed code construction \cite{jing2006distributed,chakrabarti2007low,li2006distributed}, etc. We notice, however, there are few works focusing on relay network detection and decoding \cite{yi2007joint,liu2007minimum,huang2007decode,wang2011joint,zeng2012joint}, especially when only partial channel state information (CSI) is available at the receiver end.
In practice, cooperating nodes often do not have the luxury of forwarding training symbols to the destination for channel estimation, whereas source node has the obligation of sending sufficient pilot signals for the receiver to obtain accurate channel information.
Hence, we will investigate the scenario in which only source-to-destination CSI is available at the receiver end, while the relay-to-destination CSI is unknown.

In spite of the much improved capacity and reliability, wireless channel fading and noise effects can still lead to substantial detection errors.
In practice, forward error correction (FEC) code, such as turbo code or LDPC code, is routinely conjunct with MIMO system. Particularly, LDPC code gains high popularity owing to its excellent error correction performance \cite{gallager1962low,mackay1999good}.
However, the highly nonlinear decoding process of LDPC code makes it challenging to integrate the decoding with detection step. But recently, Feldman \textit{et al.} \cite{feldman2005using} proposed a linear programming (LP) based LDPC decoding scheme, which is amendable to our detection algorithms.
This set of code constraints have been used in a variety of works: 
space-time coded MIMO system is concatenated with LDPC code in \cite{wang2014joint,wang2015joint},
wireless systems with partial channel information are treated in \cite{wang2015diversity,wang2016diversity},
multi-user MIMO using different channel codes or different interleaving patterns are handled in \cite{wang2016robust,wang2016fec},
joint SDR-based MIMO detector is proposed in \cite{wang2018non,wang2018iterative,wang2018integrated} that can approach ML performance,
and a class of asymmetric LDPC code is used in blind equalization \cite{wang2018semidefinite}.
Nonetheless, the number of parity check constraints associated with the decoding is exponentially many. If the LDPC code is sufficiently long or the parity check matrix has large row weights, the scale of the resulting LP would be prohibitive \cite{wang2017galois}. Therefore, we propose to use the cutting plane technique so that only useful parity check inequalities are integrated. By following this adaptive fashion, we can reduce the complexity to a great extent.

In the sequel, the system model is described first.
The second section elaborates on the basic formulations of detection without channel coding. Later on, integration of LDPC code constraints will be introduced in Section \ref{sec:ldpc_constr}. Furthermore, we will discuss the complexity reduction by cutting plane in Section \ref{sec:cut_sen}.
Numerical results of the proposed detection-decoding algorithms will be illustrated in Section \ref{sec:result}, and this work is wrapped up in Section \ref{sec:summary}.

\section{System Description} \label{sec:sys_model}
\begin{figure}[!htp]
\begin{center}
\includegraphics[scale=0.55]{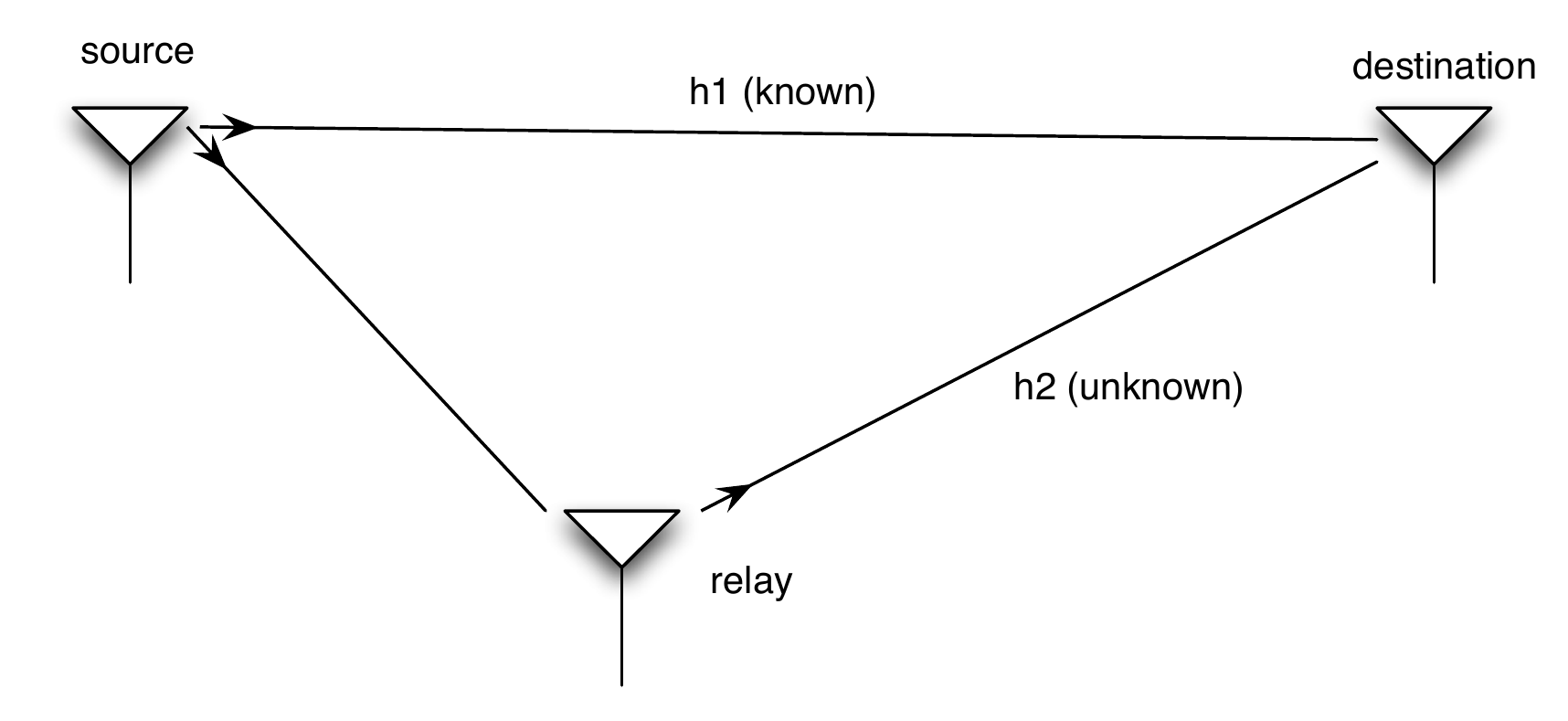}
\caption{Diagram of Relay Network with Single Relay Node.} \label{relay_diag}
\end{center}
\end{figure}

Consider the relay network shown in Fig.~\ref{relay_diag}. The source and destination have a direct channel link, and there is a relay node helping with the transmission.
Due to the limitations mentioned before, we assume that each equipment in this network is configured with a single antenna. 
In this work, we suppose the channels are Rayleigh fading, that is, $h_1$ and $h_2$ are simply zero-mean circularly symmetric complex Gaussian random variables.
In reality, $h_1$ and $h_2$ have different channel gains due to different distances. 
To incorporate this consideration, we assume $h_1 \sim \mathcal{CN} (0, \sigma_1^2)$ and $h_2 \sim \mathcal{CN} (0, \sigma_2^2)$ where $ \sigma_1^2 < \sigma_2^2$. 
A further assumption is that the channels are quasi-stationary so that they remain unchanged over a certain period of time, specially the time of transmitting two codewords in this work.

Assume in this DF relay network, there is no detection or decoding error at the relay node since the source can always find a relay node with good inter-user channel in an opportunistic manner.
Due to the consideration of power consumption, relay node does not forward pilot symbols for the estimation of relay-to-destination channel, even though the source broadcasts training symbols to it.
Moreover, assume that source-to-destination CSI is perfectly estimated at the destination by utilizing sufficient pilot symbols.
In a word, the channel $h_1$ is completely known, whereas $h_2$ is totally unknown.

Because the signals via relay node have a longer transmission and processing delay, the received signals from source and relay do not superpose at the destination. Therefore, we can collect the received signal in the following form
\begin{equation} \label{sys_eq}
\mathbf{r}[k] =
\begin{bmatrix}
r_1 [k] \\
r_2 [k]
\end{bmatrix}
=
\begin{bmatrix}
h_1 \\
h_2
\end{bmatrix}
x [k] +
\begin{bmatrix}
n_1 [k] \\
n_2 [k]
\end{bmatrix}
,\quad 1 \leq k \leq N
\end{equation}
where $x[k]$ is the transmitting signal from the source, $r_1[k]$ and $r_2[k]$ are the received signal from source and relay respectively, and $n_1[k]$ and $n_2[k]$ are additive white Gaussian noise (AWGN) at the receiver. Here $k$ denotes time slots, so totally $N$ symbols are transmitted within each quasi-stationary period.

\section{Problem Formulations} \label{sec:prob_formu}

\subsection{LP Formulation}
With the known $h_1$ in hand, we can easily detect $x[k]$ through the direct transmission by maximum likelihood (ML) or zero forcing (ZF), which is equivalent to
\begin{equation} \label{eq:obj_fun}
\underset{x[k] \in \mathbb{S}}{\text{min.}} \; \sum_{k=1}^N \vert h_1 x[k] - r_1 [k] \vert,
\end{equation}
where $\mathbb{S}$ is the set that $x[k]$ belongs to. If the set $\mathbb{S}$ is a signal constellation of finite alphabet, this minimization corresponds to ML detection; otherwise, $\mathbb{S}$ being equal to the set of complex number corresponds to ZF detection. 

To utilize the information from relay node, we introduce a combiner $\boldsymbol{\theta}$ so that received symbols from source and relay are combined to form the desired channel input $x[k]$. Mathematically, we want $x[k] = \boldsymbol{\theta}^H \mathbf{r}[k]$. However, with the presence of noise, it is desirable to have a small number $\tau[k]$ to bound the error $\vert x[k] - \boldsymbol{\theta}^H \mathbf{r}[k] \vert$. Therefore, a minimization problem can be formulated
\begin{equation} \label{eq:pcsi_lp1}
\begin{aligned}
& \underset{\boldsymbol{x}}{\text{min.}}
& &  \sum_{k=1}^N \vert h_1 x[k] - r_1 [k] \vert \\
& \text{s.t.}
& & \vert x[k] - \boldsymbol{\theta}^H \mathbf{r}[k] \vert \leq \tau [k], \quad 1 \leq k \leq N,
\end{aligned}
\end{equation}
where $\mathbf{x} = \left[ x[1] \; \ldots \; x[N] \right]$ and $\boldsymbol{\theta}$ are the variables. 

Any arbitrary predefined small number $\tau[k]$ may render the optimization problem infeasible, so we decide to consider $\tau [k]$'s as variables and lift them into the cost function to be minimized as well. 
By introducing auxiliary variable $t[k]$ to account for $\vert h_1 x[k] - r_1 [k] \vert$, we arrive at a linear program (LP) as follows
\begin{equation} \label{eq:pcsi_lp2}
\begin{aligned}
& \underset{\boldsymbol{x}}{\text{min.}}
& &  \sum_{k=1}^N (t[k] + \tau[k]) \\
& \text{s.t.}
& & \vert h_1 x[k] - r_1 [k] \vert \leq t[k], \quad 1 \leq k \leq N, \\
&
& & \vert x[k] - \boldsymbol{\theta}^H \mathbf{r}[k] \vert \leq \tau [k], \quad 1 \leq k \leq N, \\
& 
& & t[k], \tau[k] \geq 0, \quad 1 \leq k \leq N.
\end{aligned}
\end{equation}

For the problem (\ref{eq:pcsi_lp2}) being solvable by a ready-to-use solver, we need to separate the real and imaginary part of the modulated symbols. Moreover, different weights $\lambda_t$ and $\lambda_{\tau}$ are put on $t[k]$'s and $\tau[k]$'s to address the source constraints and relay constraints, respectively. Thus, we obtain the following formulation

\begin{equation} \label{eq:pcsi_lp3}
\begin{aligned}
& \underset{\boldsymbol{x}}{\text{min.}}
& &  \lambda_t \sum_{k=1}^N (t^R[k] + t^I[k]) + \lambda_{\tau} \sum_{k=1}^N (\tau^R[k] + \tau^I[k]) \\
& \text{s.t.}
& & \vert \text{Re}\{h_1 x[k]\} - \text{Re}\{r_1[k]\} \vert \leq t^R[k], \quad 1 \leq k \leq N, \\
&
& & \vert \text{Im}\{h_1 x[k]\} - \text{Im}\{r_1[k]\} \vert \leq t^I[k], \quad 1 \leq k \leq N, \\
&
& & \vert \text{Re} \{x[k]\} - \text{Re} \{ \boldsymbol{\theta}^H \mathbf{r}[k] \} \vert \leq \tau^R[k], \quad 1 \leq k \leq N, \\
&
& & \vert \text{Im} \{x[k]\} - \text{Im} \{ \boldsymbol{\theta}^H \mathbf{r}[k] \} \vert \leq \tau^I[k], \quad 1 \leq k \leq N, \\
& 
& & t^R[k], t^I[k], \tau^R[k], \tau^I[k] \geq 0, \quad 1 \leq k \leq N.
\end{aligned}
\end{equation}

Taking the finite alphabet into consideration, we can further impose integer constraints on variables $\text{Re} \{x[k]\}$ and $\text{Im} \{x[k]\}$. For example, as to the commonly used 4-QAM, we could impose $\text{Re} \{x[k]\}, \text{Im} \{x[k]\} \in \{-1,1\} $. Such constraints lead to mixed integer linear program (MILP). Alternatively, considering the fact that LP solution is always achieved at a vertex of the polyhedron defined by the constraints, we can relax the constellation constraints into box constraints $-1 \leq \text{Re} \{x[k]\}, \text{Im} \{x[k]\} \leq 1$ by hoping that these constraints are tight.  Nonetheless, this relaxation will result in performance loss, as we will see shortly in the numerical test.

\subsection{SOCP Formulation}
Due to the consideration of AWGN, the $\ell_1$-norm metric can be replaced by $\ell_2$-norm, which results in a second-order cone program (SOCP) as shown below
\begin{equation} \label{eq:pcsi_socp}
\begin{aligned}
& \underset{\boldsymbol{x}}{\text{min.}}
& &  \lambda_t \sum_{k=1}^N (t^R[k] + t^I[k]) + \lambda_{\tau} \sum_{k=1}^N (\tau^R[k] + \tau^I[k]) \\
& \text{s.t.}
& & \Vert \text{Re}\{h_1 x[k]\} - \text{Re}\{r_1[k]\} \Vert_{\ell_2} \leq t^R[k], \quad 1 \leq k \leq N, \\
&
& & \Vert \text{Im}\{h_1 x[k]\} - \text{Im}\{r_1[k]\} \Vert_{\ell_2} \leq t^I[k], \quad 1 \leq k \leq N, \\
&
& & \Vert \text{Re} \{x[k]\} - \text{Re} \{ \boldsymbol{\theta}^H \mathbf{r}[k] \} \Vert_{\ell_2} \leq \tau^R[k], \quad 1 \leq k \leq N, \\
&
& & \Vert \text{Im} \{x[k]\} - \text{Im} \{ \boldsymbol{\theta}^H \mathbf{r}[k] \} \Vert_{\ell_2} \leq \tau^I[k], \quad 1 \leq k \leq N, \\
& 
& & t^R[k], t^I[k], \tau^R[k], \tau^I[k] \geq 0, \quad 1 \leq k \leq N.
\end{aligned}
\end{equation}
Similar to MILP, we can get a mixed integer SOCP (MISOCP) by restricting the variables $\text{Re} \{x[k]\}, \text{Im} \{x[k]\} \in \{-1,1\} $, or continuous SOCP by relaxing the finite alphabet constraints to box constraints.

\section{Integration of LDPC Code Constraints}  \label{sec:ldpc_constr}
We advocate the integration of detection and decoding at the receiver in a unified optimization process. Instead of applying the transitional belief propagation between the maximum likelihood detector and the sum-product decoder in the iterative turbo equalization, we would like to incorporate a set of linear constraints that are generated from the LDPC parity check nodes.

Consider a $(N_c,K_c)$ LDPC code $\mathcal{C}$. Let $\mathcal{M}$ and $\mathcal{N}$ be the set of check nodes and variable nodes of the parity check matrix $\mathbf{H}$, respectively. That is, $\mathcal{M} = \{1, \ldots, N_c-K_c \}$ and $\mathcal{N} = \{1, \ldots, N_c \}$.
The corresponding codeword polytope can be defined as follows
\begin{equation} \label{eq:code_polytope}
\text{poly} (\mathcal{C}) = \left\{ \sum_{ \mathbf{c} \in \mathcal{C} } \lambda_{c} \mathbf{c} \; \vert \; \lambda_{c} \geq 0, \,  \sum_{ \mathbf{c} \in \mathcal{C} }  \lambda_{c} = 1  \right\}.
\end{equation}
Note that $\text{poly} (\mathcal{C})$ includes those vertices representing valid codewords, and every point in $\text{poly} (\mathcal{C})$ corresponds to a vector $\mathbf{f} = \left[f[n]\right]_{n \in \mathcal{N}}$ which is a convex combination of all the valid (integral) codewords. To describe $\text{poly} (\mathcal{C})$ for practical use, $\text{poly} (\mathcal{C})$ can be relaxed into the so-called fundamental polytope obtained by intersecting the convex hull of local codewords. Denote the set of neighbors of the $m$-th check node as $\mathcal{N}_m$. For a subset $\mathcal{F} \subseteq \mathcal{N}_m$ with odd cardinality $|\mathcal{F}|$, the explicit characterization of $\text{poly} (\mathcal{C})$ is given in \cite{feldman2005using} by the following parity check inequalities
\begin{equation} \label{eq:parity_ineq}
\sum_{ n \in \mathcal{F} } f[n] - \sum_{ n \in (\mathcal{N}_m \backslash \mathcal{F})} f[n] \leq |\mathcal{F}| - 1, \quad \forall m \in \mathcal{M}, \mathcal{F} \subseteq \mathcal{N}_m, |\mathcal{F}| \, \text{odd}
\end{equation}
and bit range constraints
\begin{equation} \label{eq:box_ineq}
f[n] \in \{0,1\} \quad \text{or} \quad 0  \leq f[n] \leq 1, \quad \forall n \in \mathcal{N}
\end{equation}
where the constraints $f[n] \in \{0,1\}$ corresponds to MILP or MISOCP, whereas $0  \leq f[n] \leq 1$ results in the relaxed LP or SOCP.
Note that the parity check inequalities (\ref{eq:parity_ineq}) are not an equivalent representation of $\text{poly} (\mathcal{C})$, but a relaxed one with both integral and fractional vertices included. Nonetheless, the parity inequalities indeed can help to tighten our solution by explicitly forbidding every bad configurations of the codewords. To see this, if parity check node $m$ fails to hold, there must be a subset of variable nodes $\mathcal{F} \subseteq \mathcal{N}_m$ of odd cardinality such that all nodes in $\mathcal{F} $ have the value 1 and all those in $\mathcal{N}_m \backslash \mathcal{F}$ have value 0. Clearly, the corresponding parity inequality would be violated.

To integrate the constraints (\ref{eq:parity_ineq}) and (\ref{eq:box_ineq}) into the problem (\ref{eq:pcsi_lp3}) and (\ref{eq:pcsi_socp}), the last step is to connect the symbol $x[k]$ and bit $f[n]$. For the 4-QAM with Gray mapping that we use in this manuscript, the relationship between $x[k]$ and $f[n]$ can be described by the following equality constraints:
\begin{subequations} \label{eq:qpsk_gray}
\begin{align}
\text{Re}\{ x[k] \} = & 1 - 2 f[2k], \quad 1 \leq k \leq N \\
\text{Im}\{ x[k] \} = & 1 - 2 f[2k-1], \quad 1 \leq k \leq N
\end{align}
\end{subequations}

\section{Unified LP and Cutting Plane}  \label{sec:cut_sen}
\subsection{Unified Linear Program}
With the aforementioned pieces, we are now in a good position to summarize the optimization problem unified with LDPC code constraints. Because the log-likelihood ratio (LLR) of each bit is available through the direct transmission, we are able to modify our objective function by adding the cost term $\sum_{n=1}^{N_c} \gamma[n] f[n]$ as in \cite{feldman2005using}, where $\gamma [n]$ denotes the LLR of bit $f[n]$. Therefore, we have the following unified LP (\textbf{ULP}).

\begin{equation} \label{eq:ULP}
\begin{aligned}
& \underset{\boldsymbol{x}}{\text{min.}}
& &  \lambda_t \sum_{k=1}^N (t^R[k] + t^I[k]) + \lambda_{\tau} \sum_{k=1}^N (\tau^R[k] + \tau^I[k]) + \sum_{n=1}^{N_c} \gamma[n] f[n] \\
& \text{s.t.}
& & \vert \text{Re}\{h_1 x[k]\} - \text{Re}\{r_1[k]\} \vert \leq t^R[k], \quad 1 \leq k \leq N \\
&
& & \vert \text{Im}\{h_1 x[k]\} - \text{Im}\{r_1[k]\} \vert \leq t^I[k], \quad 1 \leq k \leq N \\
&
& & \vert \text{Re} \{x[k]\} - \text{Re} \{ \boldsymbol{\theta}^H \mathbf{r}[k] \} \vert \leq \tau^R[k], \quad 1 \leq k \leq N \\
&
& & \vert \text{Im} \{x[k]\} - \text{Im} \{ \boldsymbol{\theta}^H \mathbf{r}[k] \} \vert \leq \tau^I[k], \quad 1 \leq k \leq N \\
&
& & \text{Re}\{ x[k] \} = 1 - 2 f[2k], \quad 1 \leq k \leq N \\
&
& & \text{Im}\{ x[k] \} = 1 - 2 f[2k-1], \quad 1 \leq k \leq N \\
&
& & \sum_{ n \in \mathcal{F} } f[n] - \sum_{ n \in (\mathcal{N}_m \backslash \mathcal{F})} f[n] \leq |\mathcal{F}| - 1, \quad \forall m \in \mathcal{M}, \mathcal{F} \subseteq \mathcal{N}_m, |\mathcal{F}| \, \text{odd} \\
&
& & f[n] \in \{0,1\} \quad \text{or} \quad 0  \leq f[n] \leq 1, \quad \forall n \in \mathcal{N}\\
&
& & t^R[k], t^I[k], \tau^R[k], \tau^I[k] \geq 0, \quad 1 \leq k \leq N.
\end{aligned}
\end{equation}
In this (MI)LP, the variables are $\text{Re}\{ x[k] \}$, $\text{Im}\{ x[k] \}$, $\text{Re}\{ \boldsymbol{\theta}\}$, $\text{Im} \{ \boldsymbol{\theta} \}$, $f[n]$ together with the auxiliary variables $t^R[k]$, $t^I[k]$, $\tau^R[k]$, $\tau^I[k]$. If we solve the MILP, the desired output is $f[n]$; otherwise, we can run the relaxed LP and then slice on the symbol $x[k]$ or directly round the bit $f[n]$.
The (MI)SOCP with LDPC code constraints is exactly in the same form as the unified LP except the $\ell_1$-norm replaced by $\ell_2$-norm. For the sake of simplicity, we do not bother to further show the formulation of unified (MI)SOCP, nor its complexity reduction as well as its numerical results henceforth.

\subsection{Complexity Reduction by Cutting Plane}
We notice that the parity check inequalities (\ref{eq:parity_ineq}) are exponentially many. Specifically, for the parity check matrix $\mathbf{H}$ with weight $d_m$ in the $m$-th row, we will have $\sum_{m \in \mathcal{M}} 2^{d_m -1}$ constraints correspondingly. Hence, we are urged to reduce the number of parity check constraints and thus reduce the overall complexity.
As introduced in \cite{taghavi2008adaptive}, the adaptive cutting plane method is shown to be effective in complexity reduction. More specifically, for our problem, the adaptive procedure works as follows

\begin{enumerate}
\item[S1] Initialize the \textbf{ULP} without parity check constraints (\ref{eq:parity_ineq}).
\item[S2] Solve the current LP to obtain the bits $f[n]$ in MILP or demodulate the symbols $x[k]$ to bits $f[n]$ by making hard decision in the relaxed LP.
\item[S3] If cuts (violated constraints) are found by substituting $\{ f[n] \}_{n \in \mathcal{N}}$ into the parity check inequalities, add them to LP and return to S2; otherwise, go to S4.
\item[S4] Output the bits and compute the bit error rate (BER).
\end{enumerate}

\section{Numerical Results}  \label{sec:result}

Throughout the simulations, we assume that source, relay and destination are all equipped with single antenna. The data symbols are 4-QAM modulated. Suppose $\sigma_1^2$ and $\sigma_2^2$ to be 0.5 and 1, respectively. Currently we do not have any preference on either kind of constraints, so we set equal weights $\lambda_t = \lambda_{\tau} = 1$.

\subsection{LP and SOCP without Code Constraints}
\begin{figure}[!htp]
\begin{center}
\includegraphics[scale=0.7]{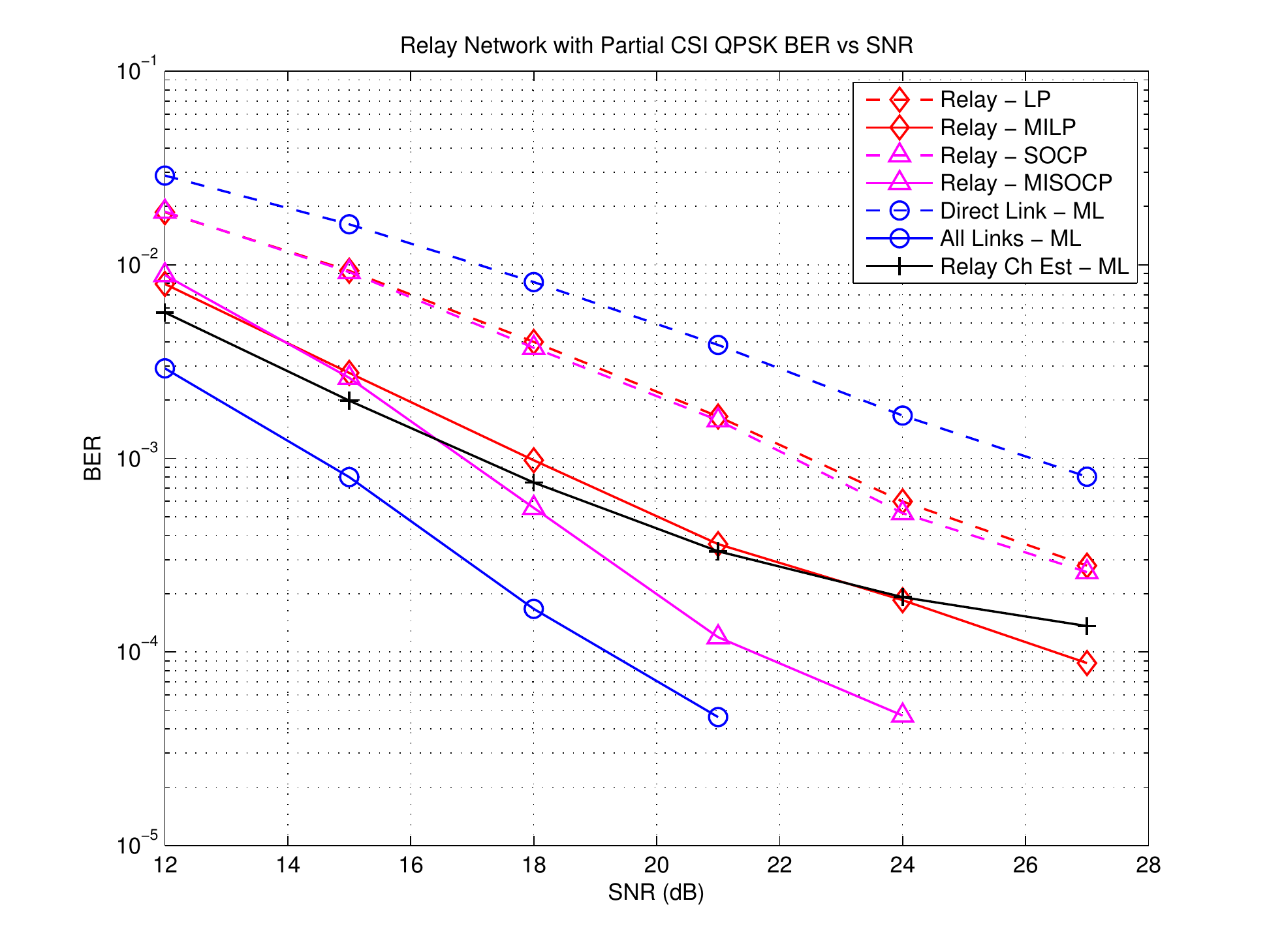}
\caption{BER Comparisons of ML, LP, MILP, SOCP and MISOCP without LDPC Code Constraints.} \label{fig:ber_comp}
\end{center}
\end{figure}

In this subsection, we test the problems in Eq.~(\ref{eq:pcsi_lp3}) and (\ref{eq:pcsi_socp}), i.e., the algorithms without LDPC code constraints.
Assume 20 QAM symbols are transmitted under one coherence time. The BER comparisons are shown in Fig.~\ref{fig:ber_comp}. The method ``Direct Link - ML" only uses the known $h_1$ for coherent ML detection, whereas ``All Links - ML" utilizes both $h_1$ and $h_2$ to detect. Clearly, we can see the big performance increment provided by spatial diversity. Hence, these two ML detections serve as the performance lower bound and upper bound, respectively. In addition, another method labeled ``Relay Ch Est - ML" uses the decision-feedback principle. It first detects symbols as ``Direct Link - ML''. Assuming the symbols are correctly detected, $h_2$ will be estimated and then the symbols are detected again by using both $h_1$ and $h_2$. 

The rest curves demonstrate our methods in which only $h_1$ is assumed to be known. ``Relay - LP" and ``Relay - SOCP" are the relaxed counterparts of ``Relay - MILP" and ``Relay - MISOCP", respectively. It is clear that relaxation on integer variables leads to substantial BER performance degradation. We notice that LP performs on par with SOCP, whereas MISOCP outperforms MILP more than 4 dB in high SNR regimes. Moreover, note that ``Relay Ch Est - ML" tends to have an error floor at high SNRs due to the decision feedback. 

\subsection{Unified LP and Complexity Reduction}
\begin{figure}[!htp]
\begin{center}
\includegraphics[scale=0.7]{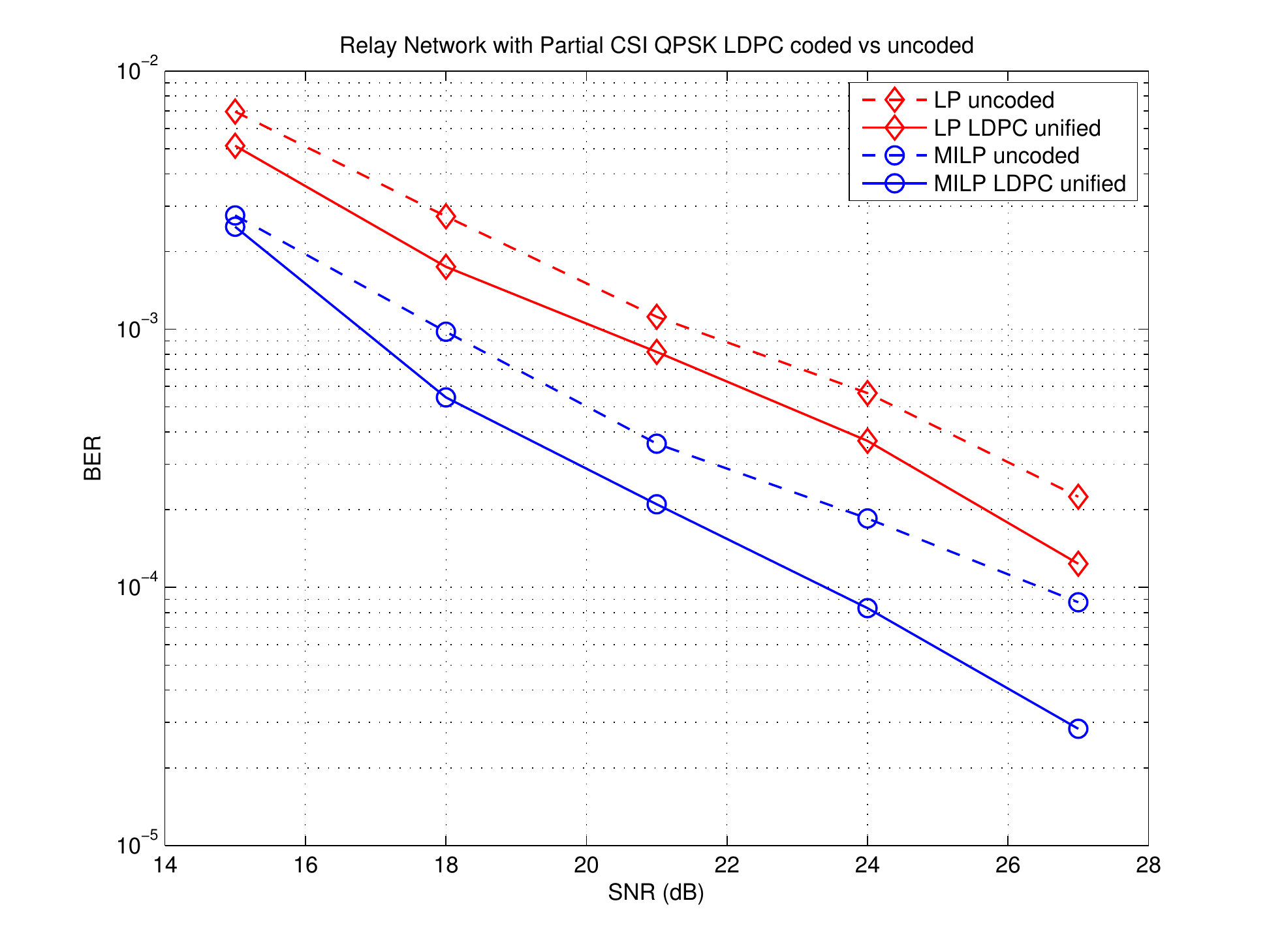}
\caption{BER Comparisons of LP and MILP with and without LDPC Code Constraints.} \label{fig:coded_uncoded}
\end{center}
\end{figure}

Firstly, we will show the performance gain that is obtained by integrating the LDPC code constraints. In particular, we only focus on LP and MILP. The LDPC code we use is a (128,96) code with code rate 3/4 and column weight 2.
The comparisons are conducted between unified (MI)LP and uncoded (MI)LP, which is shown in Fig.~\ref{fig:coded_uncoded}. As expected, the proposed unified algorithms outperform the uncoded algorithms. Specifically, MILP with LDPC code constraints has 3dB SNR gain when BER is around $10^{-4}$. In addition, we observed that this unified MILP runs much faster than the MILP without code constraints. We conjecture this is due to the LLR term in the objective function, which results in faster convergence to the optimal integer solution.

\begin{figure}[!htp]
\begin{center}
\includegraphics[scale=0.7]{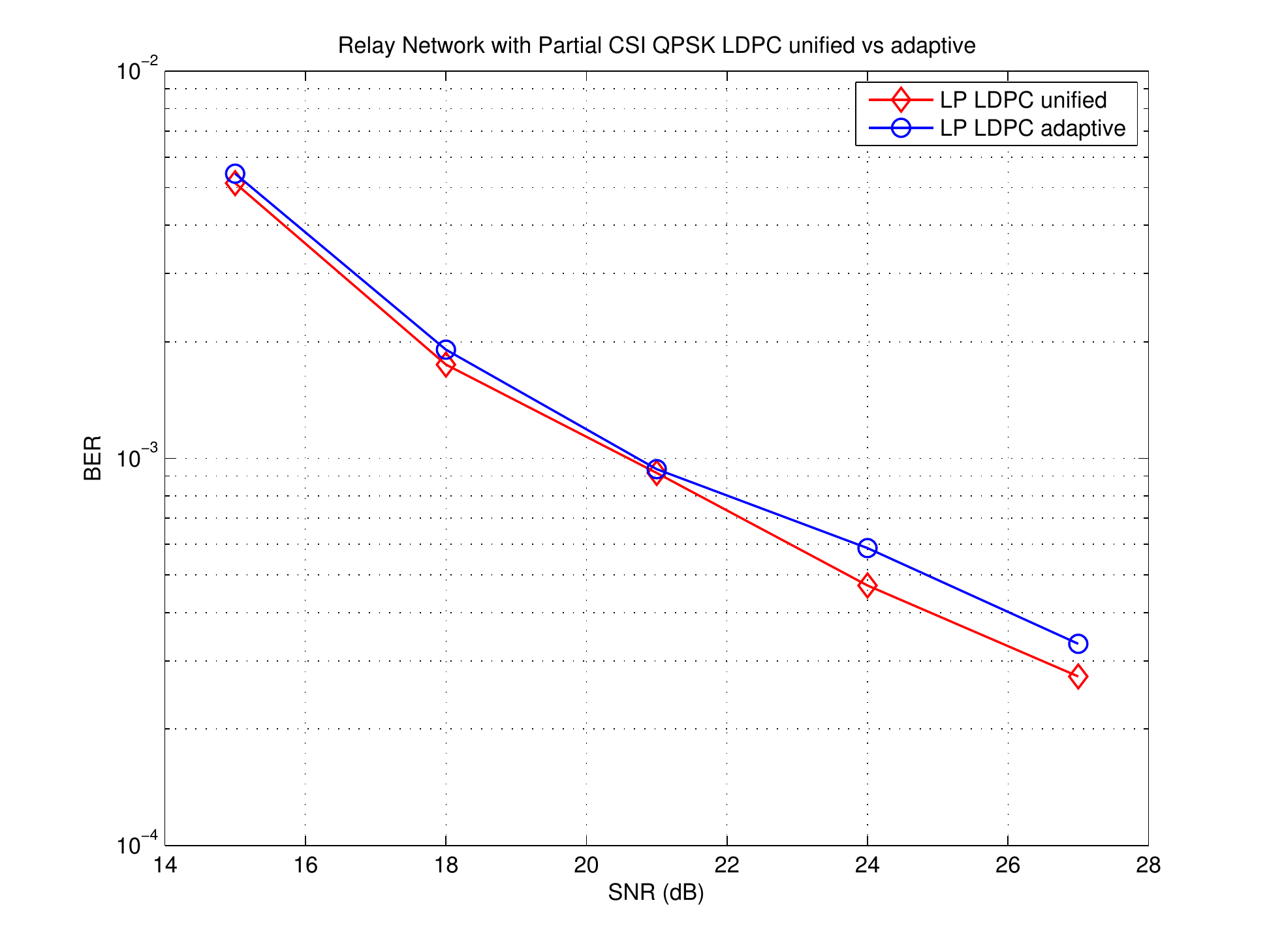}
\caption{BER Comparisons of unified LP and adaptive LP.}
\label{fig:ber_adp}
\end{center}
\end{figure}

Next we show the BER performance of the adaptive LP and the Flops comparisons of unified LP and adaptive  LP. Fig.~\ref{fig:ber_adp} illustrates the comparison between the original unified LP and adaptive LP, both with code constraints. The LDPC code used is still the (128,96) code. It is clearly seen that the adaptive LP is only a little bit inferior to the unified LP. As we will see shortly, this slight performance loss gives us a substantial complexity reduction.

\begin{figure}[!htp]
\begin{center}
\includegraphics[scale=0.7]{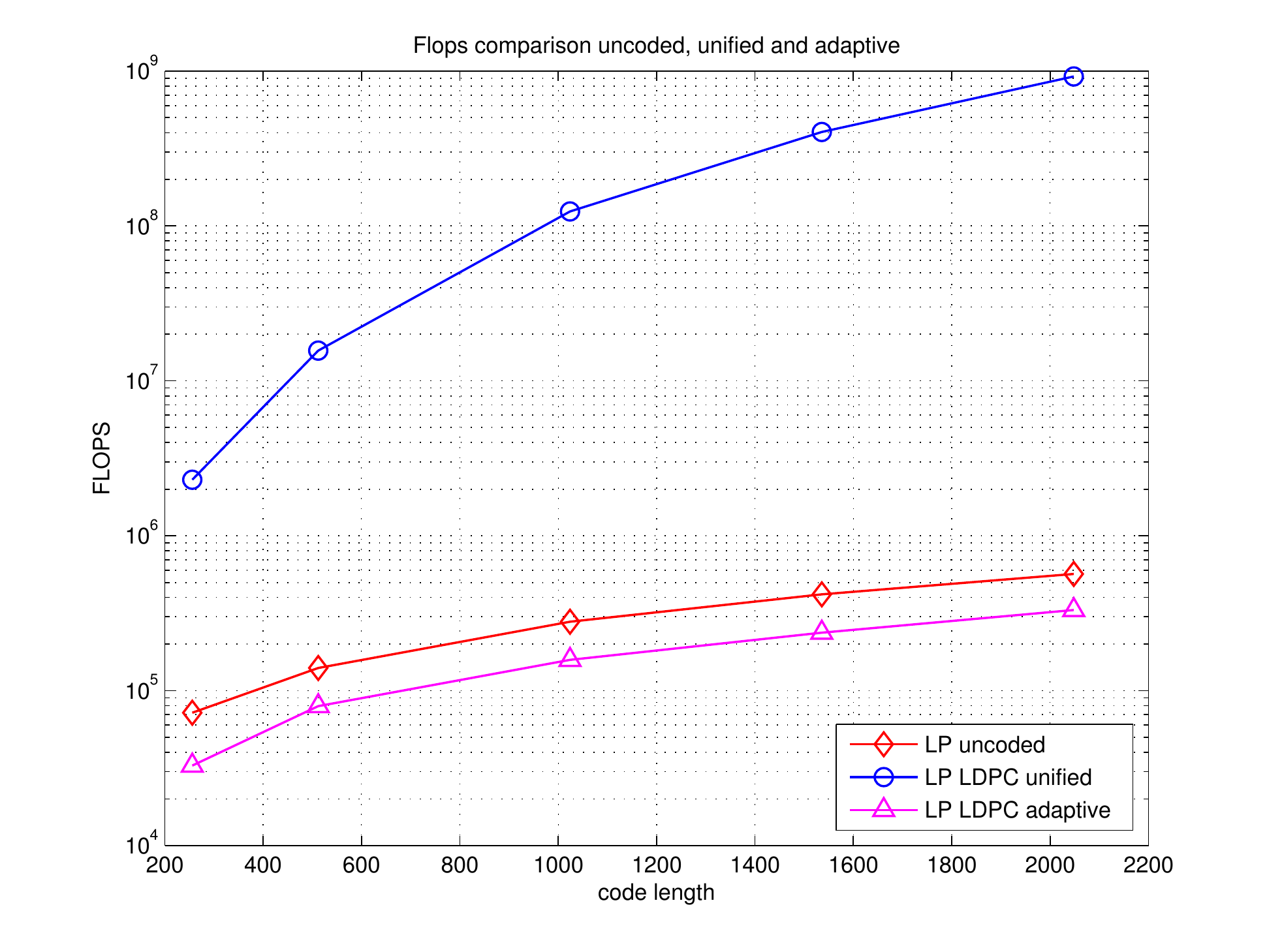}
\caption{Flops Comparisons of uncoded LP, unified LP and adaptive LP.}
\label{fig:flops}
\end{center}
\end{figure}

Now we proceed to show the great impact of adaptive LP on the reduced complexity.
Instead of using the (128,96) code only, we use a bunch of LDPC codes this time. All the tested LDPC codes are regular codes with column weight 3 and code rate 1/2. So the only factor affecting the number of code constraints is the code length. For clarity, we list the codes information in Table I. From the 2nd and 3rd row of the table, we can see the parity check inequalities constitute the majority of  total constraints in unified LP. By using cutting plane, it is observed that only a small number of parity check inequalities are added, especially in the high SNR regime, so the complexity is reduced largely. The Flops comparison, read from MOSEK, is shown in Fig.~\ref{fig:flops}. We can clearly see the amazing complexity reduction from unified LP to adaptive LP. It is also noted that adaptive LP is even better than uncoded LP in the Flops comparison, although adaptive LP has larger number of variables and constraints than uncoded LP. The reason is probably due to the LLR introduced in the cost function of unified/adaptive LP.

\begin{table}[!htp]
\centering
\label{table:code}
\caption{List of LDPC Code Information}
\begin{tabular}{|c|c|c|c|c|c|}
\hline
Code length & 256 & 512 & 1024 & 1536 & 2048 \\
\hline
\# of parity check constrs & 4432 & 8784 & 16848 & 24656 & 32832 \\
\hline
\# of total constrs in unified LP & 5712 & 11344 & 21968 & 32336 & 43072 \\
\hline \hline
\# of variables in adaptive LP & 1028 & 2052 & 4100 & 6148 & 8196 \\
\hline
\# of constrs in adaptive LP & 1280 & 2563 & 5120 & 7680 & 10240 \\
\hline
\# of variables in uncoded LP & 772 & 1540 & 3076 & 4612 & 6148 \\
\hline
\# of constrs in uncoded LP & 1024 & 2048 & 4096 & 6144 & 8192 \\
\hline
\end{tabular}
\end{table}

\newpage

\section{Summary}  \label{sec:summary}

In this work, we have formulated the basic detection problem in the form of (MI)LP and (MI)SOCP without code constraints. To tighten the solution, we propose to integrate the LDPC code constraints. However, the number of parity check inequalities is exponentially many. We further use cutting plane method to significantly reduce the complexity of the unified LP with affordable performance loss. In addition, our proposed algorithms show substantial performance gain over existing algorithms. Future works will consider the more realistic scenario with decoding error at relay nodes and address more efficient utilization of the bandwidth in the LDPC-coded relay network. It is also interesting to investigate the robust receiver's performance against RF imperfections, such as I/Q imbalance and phase noise \cite{wang2017phase}.
Moreover, joint design of precoder \cite{wu2015cooperative} and receiver would be a good topic to pursue.

\ifCLASSOPTIONcaptionsoff
  \newpage
\fi



%

\bibliographystyle{IEEEtran}
\bibliography{IEEEabrv,mybibfile}

\end{document}